# Quasi-1-Dimensional Superconductivity in Highly Disordered NbN Nanowires

K. Yu. Arutyunov[1,2], A. Ramos-Álvarez[3], A. V. Semenov[4,5], Yu. P. Korneeva[4], P.P. An[4], A. A. Korneev[1,4], A. Murphy[6], A. Bezryadin[6], and G. N. Gol'tsman[1,4]

[1]*National Research University Higher School of Economics, Moscow Institute of Electronics and Mathematics,109028, Moscow, Russia*
[2]*P.L. Kapitza Institute for Physical Problems RAS, Moscow, 119334, Russia*
[3]*LBTS, Facultad de Física, Universidade de Santiago de Compostela, ES-15782, Santiago de Compostela, Spain*
[4]*Moscow State Pedagogical University, 119991, Moscow, Russia*
[5]*Moscow Institute of Physics and Technology, 141700, Dolgoprudny, Russia*
[6]*Department of Physics, University of Illinois at Urbana-Champaign, Urbana, IL 61801-3080, USA*

The topic of superconductivity in strongly disordered materials has attracted a significant attention. In particular vivid debates are related to the subject of intrinsic spatial inhomogeneity responsible for non-BCS relation between the superconducting gap and the pairing potential. Here we report experimental study of electron transport properties of narrow *NbN* nanowires with effective cross sections of the order of the debated inhomogeneity scales. We find that conventional models based on phase slip concept provide reasonable fits for the shape of the *R(T)* transition curve. Temperature dependence of the critical current follows the text-book Ginzburg-Landau prediction for quasi-one-dimensional superconducting channel $I_c \sim (1-T/T_c)^{3/2}$. Hence, one may conclude that the intrinsic electronic inhomogeneity either does not exist in our structures, or, if exist, does not affect their resistive state properties.

Coexistence of strong disorder and superconductivity, being a macroscopically coherent state, is the very intriguing topic. Of particular interest is the superconductor-insulator transition (SIT) observed in highly disordered two-dimensional (2D) thin films[1] as well as in ultra-thin superconducting nanowires[2,3]. Though the phenomenon has been discovered more than twenty years ago, the debates about its origin are still vivid, both in relation to thin wires[4,5] as we all thin films [6,7,8,9,10].

Recent experiments[11,12,13] on three representative materials *InO$_x$*, *NbTi* and *NbN* indicate the existence of 'intrinsic electronic inhomogeneity', claimed to be not determined by chemical or/and structural imperfection of the films. While the scanning tunnel microscopy (STM) technique can indeed reveal the spatial variation of the superconducting order parameter, the corresponding electron transport measurements in 2D films cannot shed light on the internal inhomogeneity, if it is present, for the following reason: As soon as a single channel of supercurrent is formed across a 2D superconductor, it shunts all non-superconducting inclusions. Hence, in 2D geometry inhomogeneity-dependent deviations from fluctuation-governed behavior can be resolved by electron transport experiments only at the top of R(T) transition at T>T$_c$. In the opposite limit T<T$_c$ the R(T) dependencies in 2D samples are 'sharp' irrespectively of the film uniformity.

On the contrary, in quasi-1D limit with the effective diameter of the channel smaller than the superconducting coherence length $(wt)^{1/2}<\xi$, the shape of the R(T) transition is very sensitive to inhomogeneities (here *t* and *w* are the thickness and the width of the wire, respectively). In homogeneous quasi-1D superconducting system at T<T$_c$ finite resistivity can only be originated from the impact of peculiar topological singularities of the superconducting order parameter – the *phase slips*[14] – activated either by thermal, or quantum fluctuations[15,3]. Deviation of the R(T) shape from predictions of the corresponding fluctuation model is the typical indication of inhomogeneity of a 1D sample[16]. Indeed such deviations have been observed in samples which were made inhomogeneous on purpose, by means of creating surface tension through the coating of the thin superconducting wires with silicon oxide[17].

Experiments with relatively wide *NbN* channels revealed the vortex-induced resistivity to dominate over the fluctuation mechanism[18,19,20]. Analysis of electron transport data with narrower *NbN* channels $w/\xi(0)\approx 25$ indicated that under conventional experimental conditions of small bias currents $I<<4ek_BT/h$ the *thermally-activated phase slip* (TAPS) mechanism takes over the vortex scenario[21]. It should be noted that in Ref. 20 the samples with width w=100 nm did not represent truly the 1D limit. To fit the R(T) data the authors had to modify the orthodox TAPS model[22,23], justified exclusively for quasi-1D objects $max(w,t)<\xi$, into a 'phase slip strip' scenario[20].

Recent STM study[24] of 2D *NbN* films, *ex situ* fabricated using similar conditions as our samples, revealed deviations from BCS scenario: the higher the level of disorder, the more unusual are the tunneling spectra. In particular it has been found that in *NbN* there exist reproducible from sample to sample inhomogeneity characterized by the two spatial length scales. The smaller one is of the order of few nanometers, thus is close to the coherence length $\xi$. The larger one is of several tens of nanometers. The latter one has been associated with slight variation of the film thickness due to underlying atomic steps of the substrate.

The objective of this paper is to clarify the issue of the 'intrinsic inhomogeneity' of disordered *NbN* films through study of R(T) and V(I) dependencies in quasi-1D superconducting nanowires. The nanowires were fabricated using the same technological process as the 2D films[20,21]. If the intrinsic inhomogeneity affects not only the surface properties probed by STM, but also the 'bulk', then our studies should not reveal any reproducible correlation between the transport properties and the geometrical dimensions, but rather reflect the particular 'fingerprint' of inhomogeneity distribution specific for each sample.

TABLE I. Sample parameters: sample code, experimental critical temperature $T_c$ defined as $R(T_c)=0.9R_N$, normal state resistance $R_N=R(T=15K)$, normal state resistance per coherence length $R_{\xi(0)}$, nanowire width $w$, film thickness $t$, normal state resistivity $\rho_N$ and resistance per square $R_\square$. Length $L=5$ μm was the same for all samples. Mean free path $l$ can be determined from product $\rho_N l$ using literature data [25].

| Sample | $T_c$, K | $R_N$, kΩ | $R_{\xi(0)}$, Ω | $w$, nm | $t$, nm | $\rho_N$, mΩ·m | $R_\square$, Ω |
|---|---|---|---|---|---|---|---|
| 211 | 11.60 | 91.0 | 60 | 44 | 4 | 3.20 | 801 |
| 313 | 11.50 | 83.0 | 55 | 64 | 4 | 4.25 | 1062 |
| 315 | 11.25 | 39.5 | 26 | 64 | 4 | 2.02 | 506 |
| 421 | 12.80 | 32.3 | 21 | 41 | 8 | 2.12 | 265 |
| 425 | 13.30 | 21.0 | 14 | 60 | 8 | 2.02 | 252 |
| 116 | 13.00 | 14.2 | 9 | 63 | 8 | 1.43 | 179 |
| 216 | 12.10 | 65.0 | 43 | 60 | 4 | 3.12 | 780 |
| 311 | 10.95 | 45.0 | 30 | 59 | 4 | 2.11 | 528 |
| 115 | 13.50 | 14.3 | 9 | 59 | 8 | 1.35 | 169 |
| 414 | 12.95 | 26.0 | 17 | 45 | 8 | 1.87 | 234 |
| 422 | 13.45 | 28.8 | 19 | 47 | 8 | 2.17 | 271 |
| 423 | 14.10 | 24.1 | 16 | 52 | 8 | 2.01 | 251 |
| 426 | 13.30 | 20.2 | 13 | 61 | 8 | 1.97 | 246 |
| 213 | 11.55 | 95.0 | 63 | 52 | 4 | 3.95 | 988 |
| 314 | 11.05 | 42.2 | 28 | 65 | 4 | 2.19 | 549 |
| 415 | 13.40 | 22.6 | 15 | 51 | 8 | 1.84 | 231 |
| 114 | 13.00 | 16.6 | 11 | 57 | 8 | 1.51 | 189 |
| 113 | 13.20 | 18.0 | 12 | 53 | 8 | 1.52 | 190 |
| 112 | 13.10 | 20.0 | 13 | 49 | 8 | 1.58 | 197 |
| 111 | 13.05 | 22.5 | 15 | 48 | 8 | 1.71 | 214 |
| 122 | 13.60 | 21.4 | 14 | 50 | 8 | 1.71 | 214 |
| 124 | 13.90 | 26.1 | 17 | 59 | 8 | 2.46 | 308 |
| 125 | 13.10 | 20.3 | 13 | 63 | 8 | 2.05 | 256 |
| 126 | 13.40 | 19.4 | 13 | 66 | 8 | 2.03 | 254 |

We have measured 24 samples (*Table I*): *NbN* nanowires fabricated from four separate film deposition runs resulting in different normal state resistivities. The structures were patterned from *NbN* films deposited by DC reactive magnetron sputtering from *Nb* target in gas mixture of argon and nitrogen. The thickness of the film was determined by previously measured deposition rate and deposition time. The patterning was made by e-beam lithography in HSQ resist and reactive ion etching in 5:3 mixture of Ar and $SF_6$. Then the nanowires were coated by AZ1512 photoresist to prevent further oxidation of *NbN*. All structures were of the same length L=5 μm, the film thickness $t$ was either 4 or 8 nm depending on fabrication runs. The width $w$ of the line varied between ~40 nm and ~65 nm and after patterning was controlled by SEM. The normal state resistance of the nanowires $R_N$ varied from ~14 kΩ to ~95 kΩ. All samples demonstrated pronounced R(T) transitions with the experimental critical temperature varying from $T_c$=11 K to $T_c$=14 K defined as the point where the resistance drops by 10% from the normal state value: $R(T_c)=0.9R_N$.

The sample parameters (critical temperature, normal state resistivity and mean free path $l≈0.3$ nm) correlate well with literature data on thin *NbN* films[25]. Presumably in our thin *NbN* structures (4 nm and 8 nm) both the resistivity $\rho_N$, related to bulk properties, and the resistance per square $R_\square$, related to 2D properties, are equally affected by the fabrication process and are both representative to characterize the level of disorder. Note that thickness dependence of resistivity is not a common feature. For example, in another representative highly-disordered superconductor, *MoGe,* the resistivity of the material does not depend on the film thickness or the wire diameter[26]. The most probable explanation of the resistivity thickness dependence in thin *NbN* films is the existence of 1-2 nm oxidized layer on top of the film, which has been revealed by X-photoelectron spectroscopy analyses[27]. Coherence length is calculated using almost thickness independent diffusion coefficient $D=0.45$ cm$^2$/s, which can be extracted from the known dependency of the second critical field on temperature for similar 2D films[28]. Assuming the dirty limit $l<<\xi(0)$ zero temperature coherence length $\xi(0) =(\pi\hbar D/8k_BT_c)^{1/2}≈3.3$ nm and the well-known divergence $\xi(T\rightarrow T_c)$, one can consider our samples to be in quasi-1D limit $\xi(T)>max(t,w)$ within a relatively wide temperature range ~1K below $T_c$, where one can expect the phase slips (PS) mechanism(s) to be revealed. It has been shown that vortices can exist in 2D superconductors only as soon as $w>4.4\xi$ [29]. Hence, in our nanowires vortex formation can be disregarded at $T>0.9T_c$.

Structures were measured using the same electronics, which have been used previously for various sensitive experiments on mesoscopic-size superconductors[30,31]. The measuring circuit contained several stages of RLC filters capable to reduce the impact of external electromagnetic noise down to $T_{electron}$ - $T_{phonon}$ ≈ 15 mK at a base temperature $T_{phonon}≈20$ mK [32].

The R(T) dependencies were measured by using the current bias mode with the typical value $I_{ac}=10$ nA rms.

Within the experimental errors an increase of the measuring current by factor of 10 did not change the shape of the R(T) curves noticeably. There is the correlation between the experimental critical temperature $T_c$ and the sample's resistance per square $R_\square$ or resistivity $\rho_N$: The higher are these values, the lower is the critical temperature. The observation is in agreement with literature data [13,18,19,20,21,25].

The $R(T)$ data from several representative samples are presented in Figs. 1 and 2. The $R(T)$ dependencies are rather smooth and are free from obvious sample-dependent artifacts, which would immediately disable an interpretation based on trivial inhomogeneity.

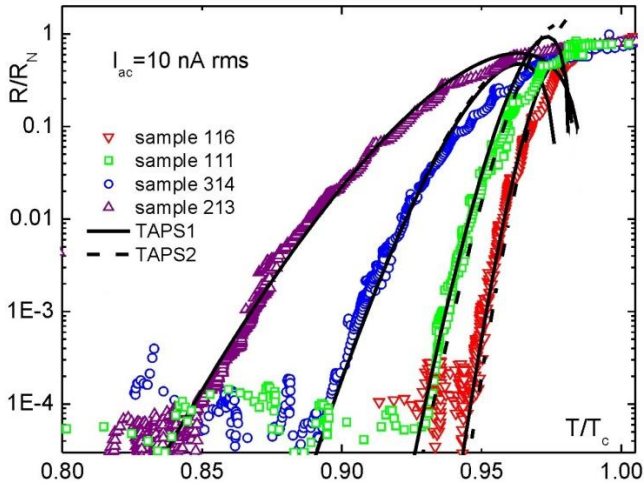

FIG. 1. Series of R(T) transitions for representative samples with low to high normal state resistivity. Normal resistances of the presented samples covers the whole range from the lowest one (sample 111, 14 k$\Omega$) to the highest one (sample 213, 95 k$\Omega$). Lines correspond to fits with TAPS model[22,23,33]. Solid lines account for the prefactor $\Omega(T)$ derived in Ref. 23, while the dashed lines for that of Ref. 33. Recording of each curve typically takes of about 30 min. Kinks on the R(T) dependencies are experimental artifacts originating from switching of the lock-in amplifier sensitivity ranges and the finite integration time ~10 s.

The straightforward interpretation of finite resistance of a homogeneous quasi-1D superconducting channel below $T_C$ evokes the fluctuation-governed phase slip process. The model of thermally activated phase slips (TAPS) [22,23,33] predicts that the resistance is governed by the activation exponent:

$$R(T) \propto \Omega(T) \exp\left(-\frac{\Delta F}{k_B T}\right). \quad (1)$$

The free energy barrier $\Delta F$ can be expressed through parameters of the nanowire as $\Delta F = 2.2\,(R_Q/R_{\xi(0)})k_B T_c (1-T/T_{c0})^{3/2}$, where $R_{\xi(0)} = R_N(L/\xi(0))$ is the sample resistance in normal state per coherence length $\xi(0)$, $R_N$ the resistance in normal state, and $R_Q = h/(2e)^2 = 6.45$ k$\Omega$ the superconducting quantum resistance. The expression for $\Delta F$ can be easily obtained within the Ginzburg-Landau (GL) model and, hence, formally is correct only rather close to the critical temperature $T_{c0}$. However in the dirty limit the quantitative discrepancy between the GL and the microscopic expressions is just about 20% at $T=0.7T_{c0}$ (see Ref. 34, Fig. 2). As in all our samples measurable resistance could be detected only at $T>0.8T_c$, hereafter we use the GL expression for $\Delta F$. The activation law Eq. (1) contains also the temperature-dependent prefactor $\Omega(T)$, which is different in Ref. [23] and in more recent work [33]. In addition it has been argued [35] that the evaluation of the 'original' prefactor [23] is correct only in the limit $\Delta(T) < k_B T$, which is not satisfied within the whole range of our measurements. However due to the relatively weak dependence of the both prefactors on temperature, compared to the exponent in Eq (1), the difference between the prefactors does not account for significant discrepancy in shape of the R(T) fits at R(T)<<$R_N$, where the phase slips are rare events and the TAPS model is essentially valid (see Fig. 1).

Fig. 1 presents R(T) dependencies for representative samples with different normal state resistances, fitted by the TAPS model Eq. (1) using prefactor $\Omega(T)$ both from Refs. 23 and 33. The only fitting parameter (besides the absolute value of the normal state resistance $R_N$) is the 'best fit' critical temperature $T_{c0}$, which turned out to be slightly lower than the empirically determined critical temperature $R(T_c)=0.9R_N$. It should be noted that the zero-temperature coherence length $\xi(0)$ is not a free parameter. One can see that the shape of $R(T)$ dependencies can be nicely fit with the TAPS model Eq. (1) for the samples with both low and high normal state resistances (Fig. 1).

At the same time, for some samples the $R(T)$ dependencies demonstrate deviations from the TAPS behavior: namely, one can observe extended low-temperature tails (Fig.2). Quantitatively, these deviations are not dramatic and can be attributed to some unidentified experimental artifacts such as finite level of external EM noise. Inhomogeneity of the samples in the form of short 'weak points' (e.g. overlooked in SEM tests narrow constrictions or inclusions of non-superconducting phase), which do not affect noticeably the normal state resistance $R_N$, can also contribute to finite TAPS-determined phase slip rate at lower temperatures. However, if to disregard those hypothetical artifacts, the deviations from the TAPS should be attributed to some other mechanism(s), which might systematically further broaden the R(T) dependencies. Below we discuss an interesting possibility that these low-temperature tails in Fig. 2 can be the manifestation of the *quantum phase slip* (QPS) mechanism [36,37].

The QPS mechanism[36,37] has been claimed to be observed in a number of experiments [15,26,38] and might be responsible for the broad R(T) transitions of the thinnest *NbN* nanowires (Fig. 2). The QPS contribution to finite resistivity of a quasi-1D superconductor is given by [15]:

$$R(T) = b \frac{\Delta(T) S_{QPS}^2 L}{\xi(T)} \exp(-2 S_{QPS}) \qquad (2)$$

where $b$ is an unimportant constant which remains the same for all samples; $\Delta(T)$ and $\xi(T)$ are the temperature dependent superconducting energy gap and coherence length, respectively. The QPS action $S_{QPS}=AR_Q/R_{\xi(T)}$, where $R_{\xi(T)}$ is the resistance in normal state per coherence length $\xi(T)$ at temperature $T$, and the constant $A\approx 1$ is the numerical prefactor that, unfortunately, cannot be determined more precisely within the model [36,37]. Sufficiently close to $T_c$, one can approximate the QPS action as $S_{QPS}=AR_Q/R_{\xi(0)}(1-T/T_{c0})^{1/2}$, similar to the expression for energy barrier $\Delta F$ in TAPS model (1). As seen from Fig. 2, fits with QPS model provide reasonable agreement with our data except the very top of the $R(T)$ transition. One should clearly understand that fitting R(T) dependencies with TAPS and QPS mechanisms, should not be understood as 'either or': both contributions should be present in all samples unless the inhomogeneities are not too severe to prohibit applicability of any model developed for uniform objects. In the thickest nanowires TAPS mechanism should dominate over the whole temperature range of experimentally measurable finite resistance. Due to exponential dependence of the QPS mechanism on cross section (Eq. 2), one can expect that in the thinnest samples at low temperature the QPS contribution can be resolved. Obviously it should be some intermediate regime, where it is difficult to separate one contribution from another. In earlier experiments on *Al* and *Ti* nanowires, where ion milling has been used to progressively reduce the cross section of a sample without affecting the bulk resistivity $\rho_N$, it has been clearly demonstrated that the crossover from TAPS to QPS mechanism is a pure size effect [15,39,40,41]. In *NbN* where the normal state resistivity $\rho_N$ and the resistance per square $R_\square$ both depend on the nanowire diameter (*Table I*), one should rather quote resistivity per unit length (see expression for the action $S_{QPS}$ in Eq. 2).

FIG. 2. Series of R(T) transitions for samples with relatively high normal state resistivity. Dashed lines correspond to TAPS model [22,23], solid lines stand for QPS model [36,37].

To make a more reliable statement about the evidence of TAPS and QPS mechanisms in our measurements, it is preferable to make a comparative quantitative analysis of the whole data set, and do not consider solely the corresponding fits to particular samples. To proceed one should note that Eqs. (1) and (2) predict different dependencies of the width of resistive transition $R(T)$ on sample resistance in normal state $R_N$. Let us define the transition width as $\Delta T_c = T_1 - T_2$, where $R_1 \equiv R(T_1)$ and $R_2 \equiv R(T_2)$ are some arbitrary chosen points within the $R(T)$ phase transition. If to define $R_N^0 \equiv R(T_{c0})$ and $R(T_{c0})/R(T_{1,2}) \equiv R_N^0/R_{1,2} \equiv C_{1,2}$, then it follows from Eq. (1) that

$$\frac{\Delta T}{T_C} \cong \left(\frac{R_{\xi(0)}}{2.2 R_Q}\right)^{\frac{2}{3}} \left\{ (\ln C_2)^{\frac{2}{3}} - (\ln C_2)^{\frac{2}{3}} \right\} \qquad (3)$$

$T_{c0}$ is not measured directly and is just a fitting parameter of TAPS model, $R_N^0$ is unknown and the relation $R_N^0/R_N$ can in principle vary from sample to sample. For instance, resistance corresponding to the best fit $T_{c0}$ for the fits presented in Fig.1 varies from $0.4R_N$ to $0.8R_N$. To avoid too sophisticated analysis, let us define $T_{c0}$ such as $R_N^0 \equiv 0.5 R_N$ for all samples. Actually such a simplification appears quite realistic following the TAPS fits with Eq. (1) and the experimental R(T) data (Fig. 1). The resistances $R_{1,2}$ should be chosen in the range of applicability of the fluctuation-governed PS models, i.e. to be small compared to $R_N$. Thus, $R_2$ was set to $3\times10^{-4}R_N$, close to instrumental zero $10^{-4}R_N$, and $R_1$ was set to $3\times10^{-2}R_N$. The resulting dependence of $\Delta T_c/T_{c0}$ on $R_N^{2/3}$ is presented in Fig. 3. The depicted analysis basically requires no fitting parameters as the variation of $R_N^0$ (and correspondingly $T_{c0}$) has very little effect due to weak logarithmic dependence.

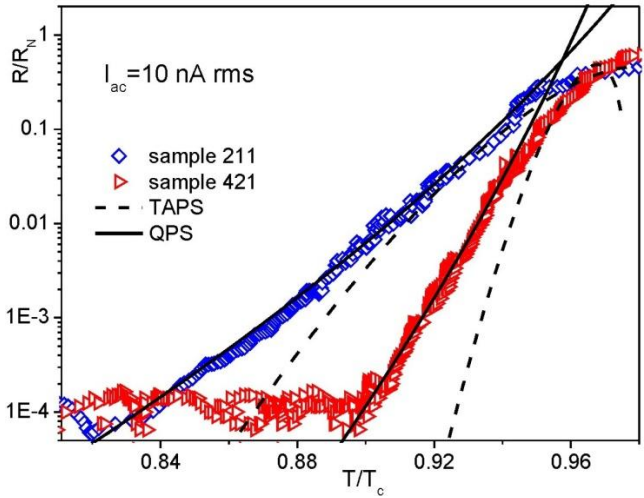

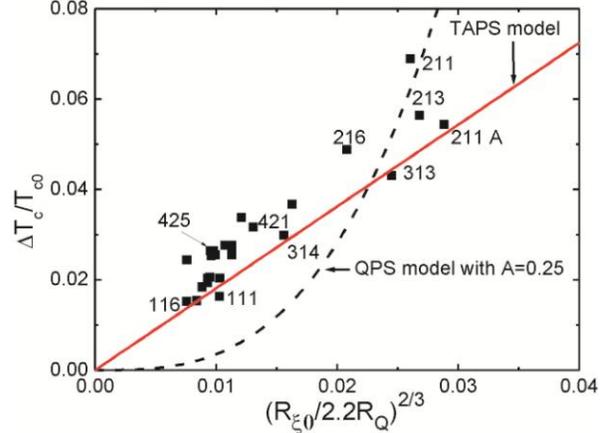

FIG. 3. Normalized width of resistive transition vs. normalized sample resistance. Points corresponding to the data presented in Fig. 1 and Fig. 2, are indicated with the specific sample codes. Point corresponding to sample 313 seems to drop out from the expected trend: resistance of the sample is two times greater than the resistances of the samples 311, 314 and 315 with close

parameters (see Table 1). Sample 211A is the same sample as 211, but cross-characterized in the other laboratory. Curves are the predictions of TAPS model (solid line) and QPS model with $A=0.25$ (dashed line).

One can notice that $\Delta T_c/T_{c0}$ for the samples with close normal state resistances has certain scattering, presumably indicating presence of inevitable sample imperfections. The transition width $\Delta T_c=T_1-T_2$ and the reference points $R_1 \equiv R(T_1)$ and $R_2 \equiv R(T_2)$ have been selected arbitrarily, and their redefinition might provide logarithmically weak better/worse agreement. Nevertheless one can conclude that almost all samples demonstrate wider transitions than TAPS model predicts.

At the same moment QPS model predicts the dependence of $\Delta T$ on $R_N$ which is qualitatively different from Eq. (3): it follows from Eq. (2) that $\Delta T_c$ should be proportional to $R_N^2$. The prediction of QPS model (Eq. (2)) is plotted in Fig 3 as the dashed line for $A=0.25$. The freedom in selection of the fitting parameter $A$ can significantly modify the QPS-governed transition width $\Delta T$. Note that in case of $Al$ and $Ti$ nanowires the best QPS fits were obtained setting $A$ to 0.15 and 0.25, correspondingly[15]. Here we found that $A=0.25\pm0.05$ provides the best fits for selected samples with the widest $R(T)$ transitions (Fig. 2). It should be noted that deviations of those high-Ohmic samples (211,213 and 216) from predictions of TAPS model is not dramatic either (Figs. 2 and 3).

The impact of coherent QPS has been observed unambiguously in qubit-type measurements in $NbN$ rings, containing narrow segments with width 30 nm or lower[42]. Rate of QPS was found to be of order $\gamma_{QPS}=10$ GHz for the nanowire with the length of 1 μm made of film with sheet resistance $R_\square=2$ kΩ which corresponds to $R_{\xi(0)}=220$ Ω if to assume that $\xi(0)$ has the same value as in our samples. Note that the obtained resistance $R_{\xi(0)}$ is more than two times greater than that of the most high-Ohmic samples studied in present work (TABLE I). Taking the directly measured data for $\gamma_{QPS}$ from Ref. 42 as the reference, it is possible to estimate the expected effective resistance due to QPS process in our samples. One can rewrite Eq.(2) as

$$R(T) \cong \frac{L}{\xi(T)} R_Q \frac{h^2 \gamma'^2_{QPS}}{k_B T \Delta(T)} \quad (4)$$

where the QPS rate is (scaled to a segment of length $\xi(T)$) $\gamma'_{QPS} \sim \Delta(T) S_{QPS} \exp(-S_{QPS}) \equiv \gamma_{QPS}(\xi(T)/L)$, and set $A=0.25$. For the parameters of sample 211 and temperature 0.85 $T_c$, the resistance given by (4) turns out to be of order $10^{-5}$ Ω, which is far below the instrumental zero. The observation indicates that the samples studied in this work are still too thick (or low-Ohmic) to claim the clear impact of QPS mechanism on the shape of $R(T)$ transition. For majority of samples TAPS mechanism dominates within the whole range of experimentally observed finite resistance below $T_c$.

The current vs. voltage I-V dependencies, measured at temperatures $T<T_c$ are typical for reasonably homogeneous superconducting nanowires and qualitatively resemble the behavior of the shunted superconducting tunnel junctions measured at a finite temperature: At small bias the instrumental zero voltage state is observed, which smoothly approaches normal (Ohmic) state at higher currents (Fig. 4, inset). Absence of artifacts (e.g. kinks) indicates the absence of pronounced structural inhomogeneities such as regions with different values of local critical current density $j_c$ and/or sections with considerably different effective cross section $(tw)^{1/2}$. The temperature dependence of the critical current nicely follows the $I_c=I_{c0}(1-T/T_c)^{3/2}$ dependence (Fig. 4), following the prediction of the Ginzburg - Landau (GL) model for a 1D superconducting channel, where $I_{c0}$ is the critical current at $T=0$. To outline the distinction between the behavior of the thinnest (sample 211, largest $\Delta T_c$) and the thickest (sample 116, smallest $\Delta T_c$) nanowires, in inset to Fig. 4 we plot the I-V dependencies measured at the same effective temperature $T^*$ defined as $R(T^*)/R_N=10^{-3}$. One can clearly see that the thickest structure (116) demonstrates the instrumental zero voltage state over much wider current range (in units $I_c/I_{c0}$) compared to the thinnest nanowire (211). At much lower temperatures $T << T^*$ both structures demonstrate the I-V dependencies with extended current range corresponding to the zero voltage state, indicating that they are in 'truly' superconducting state in a sense that the rate of PSs is too small to provide a measurable voltage at a finite bias current.

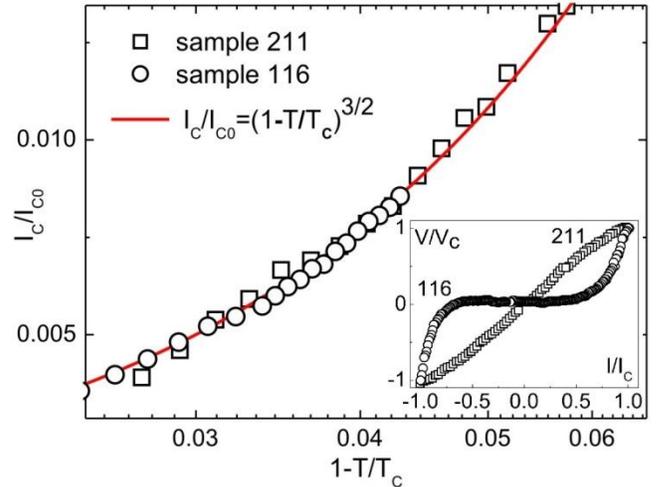

FIG. 4. Critical current vs. temperature for two representative structures with low (○, sample 116) and high (□, sample 211) normal state resistance. Line corresponds to Ginzburg - Landau $I_c=I_{c0}(1-T/T_c)^{3/2}$ dependence typical for 1D channels close to critical temperature. Inset: normalized I-V characteristics for both samples measured at temperature T*, defined as $R(T^*)/R_N=10^{-3}$. The experimental critical current $I_c$ corresponds to instrumental critical voltage $V_c=2$ μV.

As all samples demonstrate a finite measurable resistance within a certain temperature region below the critical temperature, the definition of the critical current is rather arbitrary. In the analyses above we have defined the 'critical current' as the state corresponding to the 'critical' voltage $V_c$=2 μV. This value has been selected to be well below the voltage across the sample in normal state ($T>T_c$), but noticeably larger than the experimental dc voltage zero ~10 nV to distinguish if the resistive state is approached smoothly (Fig. 4, inset) or jump-like typical for massive superconductors.

To summarize, the smooth $R(T)$ and $V(I)$ dependencies, observed for all studied samples, support our hypothesis that the intrinsic electronic inhomogeneity, revealed in STM studies of *NbN* films [13,24], either does not exist in our structures, or, if exist, does not affect their resistive state properties. Though the parameters of our most high-Ohmic samples correspond to the ones, where noticeable deviations from BCS behavior have been observed[24], we find that conventional BCS-based models of fluctuation-governed resistive state provide satisfactory agreement with our $R(T)$ data

The authors would like to acknowledge L. Kuzmin, L. Ioffe, and A. Zaikin for valuable discussions. The research carried out in 2015-2016 by K. Yu. Arutyunov was supported within the framework of the Academic Fund Program at the National Research University Higher School of Economics (HSE) in 2015- 2016 (grant No.15-01-0153) and supported within the framework of a subsidy granted to the HSE by the Government of the Russian Federation for the implementation of the Global Competitiveness Program. A. Ramos-Alvarez acknowledges financial support from Spain's MICINN through the FPI grant (no. BES-2011–046820).

[*]karutyunov@hse.ru